\PassOptionsToPackage{table,xcdraw}{xcolor}
\documentclass[biblatex,english]{lni}

\usepackage{graphicx}
\usepackage{float}
\usepackage{todonotes}
\usepackage{booktabs} 
\usepackage{siunitx} 
\usepackage{amsmath} 
\usepackage{listings}
\usepackage{stfloats}
\usepackage{censor}
\usepackage{wrapfig}
\usepackage{subcaption}
\usepackage{enumitem}
\usepackage{caption}

\begin{document}

\title[Ein Kurztitel]{Performance Analysis in Parallel Programming Education:\\ A Comparative Usability Study}
\author[1]{Anna-Lena Roth}{anna-lena.roth@cs.hs-fulda.de}{0000-0002-6463-3486}
\author[1]{David James}{david.james@cs.hs-fulda.de}{0009-0009-0625-6297}
\author[2]{Jonas Posner}{jonas.posner@uni-kassel.de}{0000-0002-6491-1626}
\author[3]{Michael Kuhn}{michael.kuhn@ovgu.de}{0000-0001-8167-8574}

\affil[1]{Fulda University of Applied Sciences\\Applied Computer Science\\Leipziger Str. 123\\36037 Fulda\\Germany}
\affil[2]{University of Kassel\\Faculty of Electrical Engineering and Computer Science\\Wilhelmshöher Allee 71--73\\34121 Kassel\\Germany}
\affil[3]{Otto von Guericke University Magdeburg\\Faculty of Computer Science\\Universitätsplatz 2\\39106 Magdeburg\\Germany}
\maketitle

\begin{abstract}
Parallel programming curricula encompass not only the development of parallel code and algorithm design but also emphasize efficiency, optimization, and performance analysis.
To equip students with the skills necessary for writing efficient parallel code using message passing with MPI, practical experience on HPC environments is essential.
Performance analysis tools assist in identifying issues such as load imbalances or bottlenecks.
Despite their use by experienced developers, these tools' complexity and required knowledge of cluster architectures, resource management, MPI, and common parallel issues hinder their educational integration.\\
To address these barriers, we developed EduMPI, a learning support tool designed to simplify cluster usage and performance analysis for students.
EduMPI offers an intuitive GUI that automates program execution on clusters and delivers near-real-time visualizations of MPI communication.
This enables students to track process communication according to their physical placement within the cluster and detect performance problems interactively.
This paper presents a user study comparing EduMPI with established professional performance analysis tools, demonstrating that EduMPI lowers entry barriers and fosters an intuitive understanding of parallel program performance, thereby enhancing its educational value.
\end{abstract}

\begin{keywords}
Parallel Programming \and MPI \and Performance Analysis \and Education \and Usability Study
\end{keywords}

\section{Introduction}
In parallel programming education, it is essential not only to teach students how to write parallel code but also to provide practical experience with real high-performance computing (HPC) environments and to develop efficient code that makes optimal use of system resources.
To achieve this, concepts for avoiding performance problems, such as load imbalances, bottlenecks, and excessive communication overhead, must be addressed.
Performance analysis is central to identifying such inefficiencies.
Several working groups behind the European Master for HPC Curriculum~\cite{BouvryBrorssonCanal2025}, the NSF/IEEE-TCPP Curriculum Initiative~\cite{PrasadEstradaGhafor2020}, and the ITiCSE Reports~\cite{RajRomanowski2020} recommend making performance analysis and visualization an integral part of parallel programming curricula.\\
Realization of this recommendation is challenging due to a lack of dedicated HPC practice environments and educational tools~\cite{RajRomanowski2020,VanDeVanterPostZosel2005}.
For many students, parallel programming courses are their first significant exposure to threads, processes, and cluster computing.
Learning communication standards such as the Message Passing Interface (MPI) is already demanding.
Performance analysis requires knowledge of code instrumentation and measurement, using appropriate tools such as Score-P~\cite{KnuepferRoessel2012} and Scalasca~\cite{GeimerWolfWylieAbrahamBeckerMohr2010}.
Most performance analysis tools work \emph{post-mortem}, collecting metrics at runtime for visual analysis after program termination.
Measurement must be repeated after every code change, causing long delays, especially on small educational clusters.
The complexity and depth of detail involved in the subsequent visual analysis with tools such as CUBE~\cite{CubeV4}, TAU~\cite{ShendeMalony2006}, or Vampir~\cite{KnuepferBrunst2008} can be overwhelming for students, as they often require a comprehensive interpretation of multiple views and metrics.
Without in-depth prior knowledge, performance issues often remain hidden initially.
This makes performance analysis in an educational environment considerably difficult.\\
At Fulda University of Applied Sciences, the course Parallel Programming was theory-heavy due to the lack of access to real HPC environments.
The introduction of an educational cluster posed initial challenges, as many students lacked prior knowledge of SSH, Bash, and SLURM~\cite{YooJette2002}, requiring substantial support.
Performance analysis was often limited to simple runtime comparisons, since professional tools were perceived as complex~\cite{RothJames2024}.\\
To address this, we developed EduMPI~\cite{RothJames2025}, a GUI-based learning tool that simplifies cluster access and, unlike common professional tools, provides near-real-time, intuitive visualizations of MPI communication without manual instrumentation or measurement.
Performance data is collected and displayed automatically during program execution, enabling seamless student engagement with performance analysis.\\
In this work, we conduct a controlled laboratory study to evaluate EduMPI's impact on students' ability to interpret runtime data and identify performance issues, as well as its usability compared to the tools CUBE and TAU.
Results show that EduMPI lowers barriers to performance analysis and fosters an intuitive understanding of parallel programs.


\section{Related Work}
In a previous study conducted during the winter semester 2023/24, we investigated to what extent the automation and accessibility of EduMPI motivate students to conduct performance analyses~\cite{RothJames2024}.
The focus was on the measurement process, comparing the user experience of EduMPI with the conventional approach of running performance measurements via the terminal using Score-P and Scalasca.
Results showed that students working with EduMPI were more motivated to perform multiple measurements, compare results, required less support, and worked more efficiently overall.
The study in this work shifts the focus to the visualization and interpretation of performance analysis results.\\
Several studies highlight the importance of real HPC environments and practical analysis in education~\cite{JoinerGrayMurphyPeck2006, YaziciMishraKarakayaZiya2016, BernreutherBrenk2005}, and others explore the integration of professional tools into teaching~\cite{DelistavrouMargaritis2010, Malakar2019}.
However, specialized educational tools for parallel programming are still rare~\cite{RajRomanowski2020}.
Zhang \textsl{et al.} developed a web-based tool for simplified parallel execution and measurement~\cite{ZhangLiWuDu2018}, while Oden \textsl{et al.} offer easy analysis workflows within Jupyter Notebooks~\cite{OdenNoelp2024}.
Atala \textsl{et al.} introduced a visualization tool for collective MPI algorithms aimed at students, though it relies on simulations rather than real measurements~\cite{AtalaMorrisonBallard2025}.
All these tools follow a post-mortem approach, showing performance data only after termination, while EduMPI offers near-real-time, beginner-friendly visualizations bridging the gap between professional complexity and educational accessibility.

\section{EduMPI}
EduMPI consists of three core components:
EduMPI GUI can be used as an AppImage without requiring installation.
MPI programs can be executed automatically on the cluster through the GUI, and MPI communication data can be interactively analyzed in near-real-time.
On the cluster side, EduMPI is installed as a fork of Open MPI~\cite{OpenMPI} and, in addition to standard Open MPI functionality, includes a data measurement system that collects MPI event-based communication data.
This data is immediately processed by the third component, a time-series database, which enables fast queries, even when handling large volumes of data simultaneously.
A detailed description of EduMPI's functionality and GUI visualization features is provided in~\cite{RothJames2025}.\\
EduMPI was developed to simplify the visualization of MPI communication in educational use.
It reduces the abstraction needed to understand process distribution across nodes/cores and their communication.
Multiple students can use EduMPI simultaneously, are exposed to performance data from the start, can identify errors or bottlenecks early, and cancel programs via the GUI to avoid blocking cluster resources.
Through custom data collection in a dedicated Open MPI fork, internal MPI data, such as point-to-point communication behind collectives, can be visualized.
This benefits both teaching and experienced developers by enabling insight into collective communication algorithms~\cite{RothJamesKuhnFrisch2025}.
Overall, EduMPI supports MPI understanding on real clusters, helps identify issues such as bottlenecks or late senders/receivers, and serves as an educational entry point to parallel performance analysis.

\section{Usability Study}
We conducted a lab-based usability study in the summer semester 2025 to evaluate how well EduMPI supports students in identifying performance problems in MPI programs and how it compares to the professional tools TAU and CUBE, both of which are open-source, freely available, and were used as exemplary tools in the course.
The study used a counterbalanced crossover design with two conditions: In version A, participants first used EduMPI, then chose to use either TAU or CUBE.
In version~B, the order was reversed.
This design allowed investigation of whether tool order affects analysis success.
Participants worked in dyads or triads at a shared workstation, with one operating the tools while the others discussed findings.
This collaborative setup was based on prior research showing that group-based testing encourages more natural verbalization of thought processes~\cite{RothJames2024}.
Sessions were fully moderated, recorded, and followed a standardized protocol.
Participants rated features and described what they liked, disliked, or found easy or difficult.
Pre- and post-session questionnaires captured both subjective (e.g., satisfaction) and objective (e.g., task correctness) usability metrics.
Protocols, questionnaires, and collected data are freely available at~\cite{RothData2025}.
In total, 33 Master's students from four study programs (Applied Computer Science, Data Science, Global Software Development, and Embedded Systems) took part in the study.
All were enrolled in the course \emph{Parallel Programming}~\cite{HSFuldaGSD}.
Participants were between 20 and 30 years old (30 male, 3 female), distributed across different stages of their studies (17 in the 1st semester, 9 in the 2nd, and 7 in later semesters), and represented seven nationalities, including 36\% Indian, 24\% German, and 18\% Pakistani.
Participants self-rated their programming proficiency as 6\% beginner, 55\% intermediate, 36\% advanced, and 3\% expert.
Regarding parallel programming, 3\% reported being slightly, 73\% moderately, and 24\% very familiar.
All tools used in the study had been previously introduced in the course.
Performance analysis workflows using Score-P and Scalasca were demonstrated in lectures, and students were given brief introductions to TAU and CUBE.
EduMPI had been used in two exercises involving structured usability tasks and giving feedback.
However, none of the tools had previously been demonstrated by teachers for identifying performance issues.
94\% of participants indicated they had prior experience with EduMPI, while 73\% had experience with TAU, CUBE, and Score-P.
This discrepancy is likely due to the voluntary nature of lectures and exercises.
In each group, however, at least one student reported familiarity with one of the tools.\\
The program examined was a parallel row-wise matrix multiplication~\cite{RothData2025}.
It follows a master-worker pattern, a structure often intuitively chosen by beginners, but is not ideal to parallelize programs~\cite{KieferWarzelTichy2015}.
The master process generates two matrices (A and B) and then distributes the entire matrix B to all processes.
Matrix A is then sent row by row to the processes, which calculate part of the result matrix C by multiplying a row of matrix A with the entire matrix B and send this back to the master, which finally combines the results.
The program was discussed in class, but without accompanying performance analysis.
This ensured focus on visual output and runtime behavior, not source code, during the usability study.
73\% of participants stated that they had attended the corresponding lecture.

\subsection{Scenario and Tasks}
Depending on the version, the program was executed with 256 processes, first using EduMPI and then Score-P, followed by analysis in CUBE or TAU, or vice versa.
As the study did not focus on the instrumentation and measurement process, EduMPI offered automated profile generation with Score-P.
Participants could choose this feature or use the traditional workflow.
Both answer correctness and insight generation were evaluated, emphasizing features that supported or hindered analysis, along with the transparency and efficiency of exploration.
The following tasks, based on core usability principles, were presented.

\begin{enumerate}[label=T\arabic*]
    \item \textbf{Process distribution across cluster nodes:}
    How and across how many nodes were the MPI processes distributed on the cluster?
    \item \textbf{Interpretation of color coding:}
    What do the different colors used to display the performance data in the respective tools mean?
    \item \textbf{Duration of the distribution of matrix B:}
    How long does it take to fully distribute matrix B, and which MPI function is used for this?
    \item \textbf{Late broadcast and bottleneck due to serial initialization:}
    Analyze the first program phase (until matrix~B is fully distributed).
    What problems arise and why?
    \item \textbf{Row-wise distribution of matrix A and calculation:}
    Examine the second program phase, where matrix A is distributed row by row to processes that compute partial results and send them back to the master.
    What problems can be identified, and why?
    \item \textbf{Centralized communication (efficiency, scalability):}
    Would you describe the MPI communication in the program as centralized or distributed?
    How do you rate the efficiency and scalability of this parallel matrix multiplication?
    \item \textbf{Dominant MPI communication (over the total runtime):}
    Consider the program over its total runtime: Which MPI function takes the most time, causes the biggest communication overhead, and transfers the largest amount of data?
\end{enumerate}

\begin{figure}
    \centering
    \vspace*{-10pt}

    \includegraphics[width=0.9\textwidth]{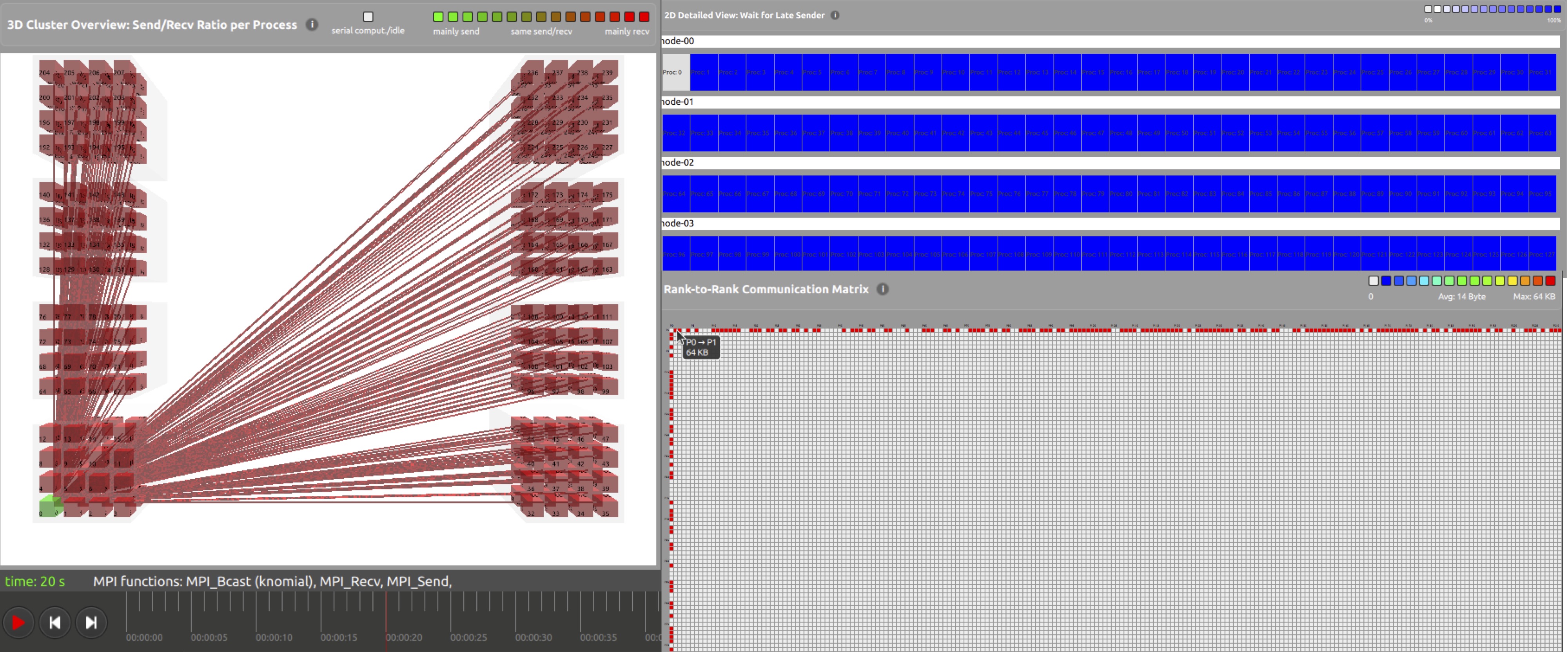}
    \captionof{figure}{EduMPI: 3D view (left) and communication matrix (bottom right) showing the second program phase; 2D view (top right) showing the first phase.}
    \label{fig:edumpi_example}

    \vspace*{10pt}

    \includegraphics[width=0.85\textwidth]{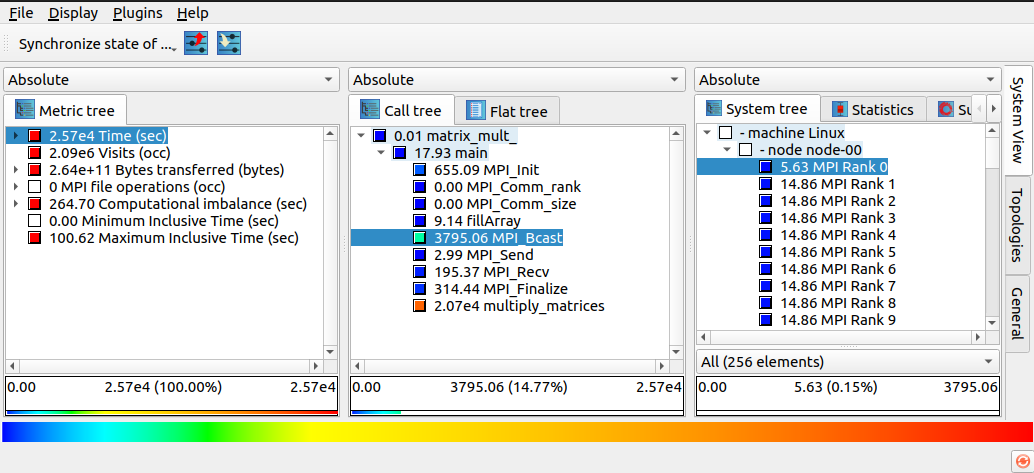}
    \captionof{figure}{CUBE: Three-pane interface with metric tree, call tree, and system tree}
    \label{fig:cube_example}

    \vspace*{10pt}

    \includegraphics[width=0.85\textwidth]{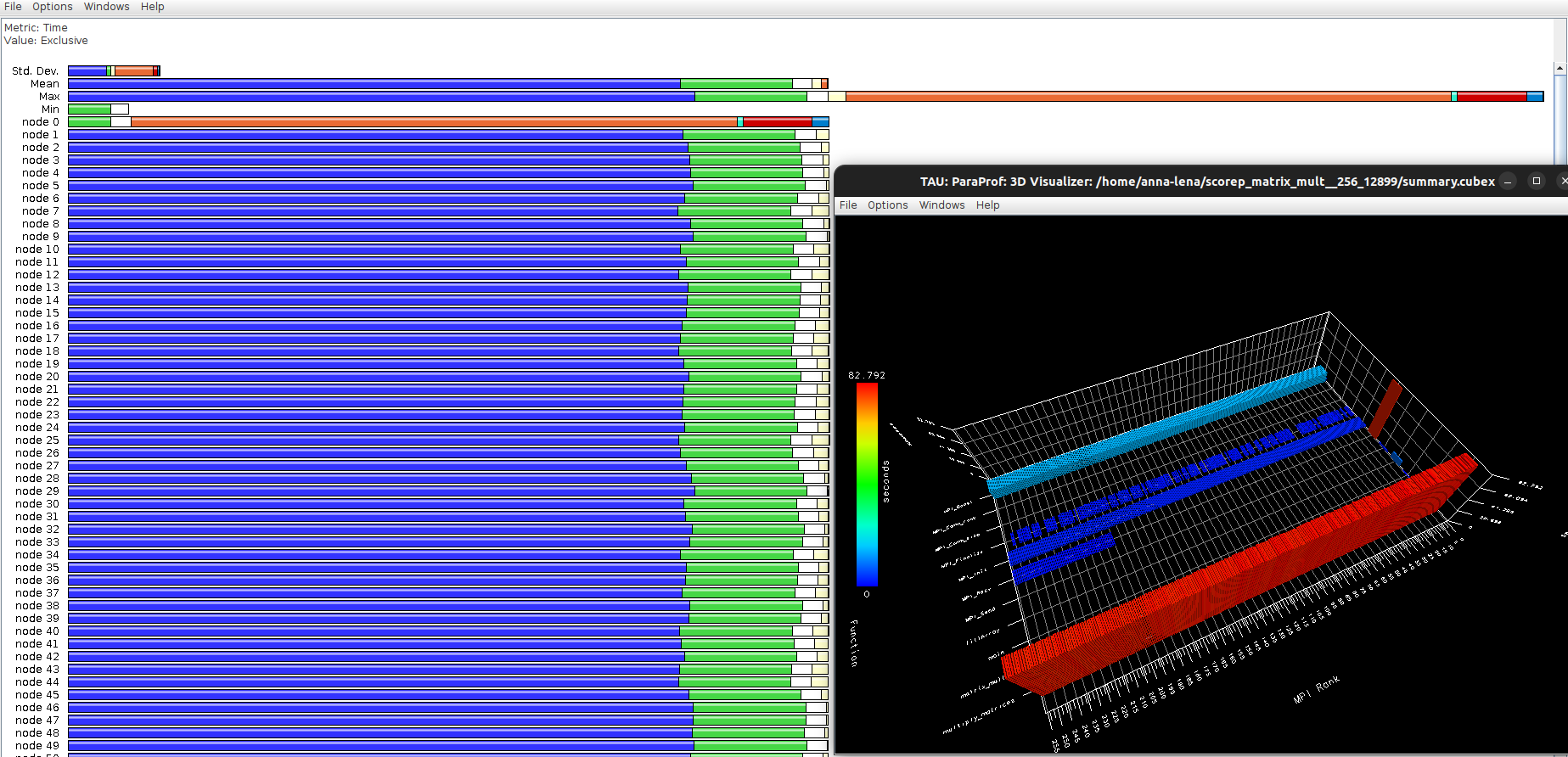}
    \captionof{figure}{TAU: 2D bar chart (left) and 3D visualization (right) of aggregated profiling data}
    \label{fig:tau_example}
\end{figure}

Figures~\ref{fig:edumpi_example}--\ref{fig:tau_example} illustrate the visualizations.
Except for T1, all tasks and issues could be addressed with each of the three tools.
EduMPI (see Fig.~\ref{fig:edumpi_example}) offers three distinct visual perspectives: a 3D view (left), a 2D view (top right), and a communication matrix (bottom right).
In the 3D view, nodes are represented as outer cubes, with their processes displayed as inner cubes.
Communication lines indicate which processes are interacting at a given point in the timeline.
The 2D view presents nodes and processes in a tabular layout, displaying metrics such as sent and received byte counts.
The communication matrix visualizes process-to-process communication via a grid, enabling detailed analysis of communication.\\ 
CUBE (see Fig.~\ref{fig:cube_example}) provides a three-pane interface: the metric tree, call tree, and system tree.
Metrics such as execution time or transferred data can be selected and analyzed.
The call tree lists executed functions along with aggregated data, while the system tree enables fine-grained analysis per process, cross-referencing metrics and function calls.
A color scale, ranging from blue (low metric values) to red (high values), supports visual interpretation.\\
TAU (see Fig.~\ref{fig:tau_example}) presents data in 2D visualizations, using a bar chart (left), and selected metrics can also be visualized in 3D plots (right).
Function-level metrics such as execution time or communication volume are displayed per process.
Multiple plots with different metrics can be generated. For example, a 3D bar plot with functions on the z-axis, MPI processes on the x-axis, and execution time on the y-axis.\\
In the first program phase, process~0 initializes the matrices while all other processes wait in an \texttt{MPI\_Bcast} call for receiving matrix B.
As initializing large matrices can take several seconds, process~0 becomes a bottleneck, and a late broadcast problem occurs.
In CUBE, this appears as process~0 spending significantly less time in the \texttt{MPI\_Bcast} call than the others (see Fig.~\ref{fig:cube_example}).
Additionally, function \texttt{fillArray}, invoked only by process~0, accounts for this time difference.
In TAU, the same pattern is reflected in the 2D bar chart: \texttt{MPI\_Bcast} appears in green, while function \texttt{fillArray} is represented in red (see Fig.~\ref{fig:tau_example}).
EduMPI includes a dedicated \emph{Wait for Late Sender} option.
Processes are colored from white to blue, reflecting the time spent blocked waiting for a sender (0–100\%).
As shown in Fig.~\ref{fig:edumpi_example} (top right), during the broadcast, all processes wait for process~0.
\texttt{MPI\_Bcast} also accounts for the largest communication volume in the program.
In the second phase, process~0 sends rows of matrix A to all other processes, which compute partial results of matrix C and send them back before receiving a next row.
All communication is routed through process~0, which can create a potential bottleneck and limit the scalability.
Since calculation time is similar across processes, delays occur when all processes simultaneously await new data.
This centralized pattern is visible in EduMPI's 3D view, when all processes exit the initial broadcast and wait for their first row of matrix A (see Fig.~\ref{fig:edumpi_example}~(left)).
The communication matrix shows all communication flows through process~0.
In both CUBE and TAU, it is visible that most processes spend the majority of their time in \texttt{multiplyMatrices}, which process~0 does not enter.
Instead, it spends a significant portion of its time in \texttt{MPI\_Recv}, indicating that it is frequently blocked, waiting to receive partial results of matrix C.

\subsection{Results}
Quantitative data (processing time, success rates, and moderation interventions) and qualitative observations (group discussions, user behavior, and questionnaires) were triangulated for analysis.
The study evaluated effectiveness, efficiency, and satisfaction across twelve sessions with groups of two or three participants.
Six groups used version A (EduMPI first), six version B (TAU or CUBE first), with nine groups choosing CUBE and three TAU.\\
\begin{figure}[b]
    \centering
    \vspace*{-10pt}
    \includegraphics[width=0.85\textwidth]{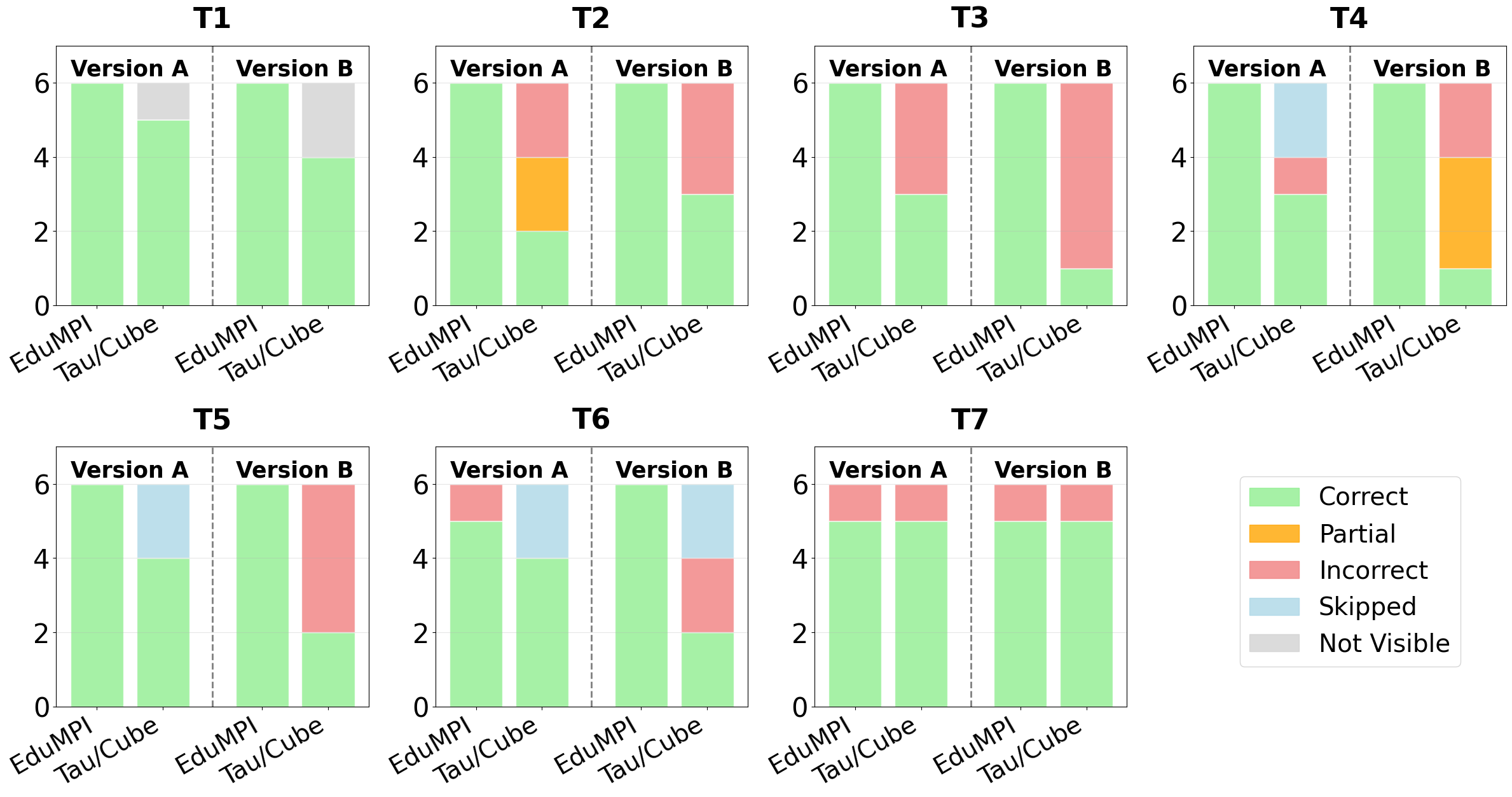}  
    \caption{Comparison of task correctness between versions and tools}
    \label{fig:correctness}
\end{figure}
\textbf{Effectiveness} was measured by the number of correctly completed tasks, whether solved independently or with moderator support (see Fig.~\ref{fig:correctness}).
With EduMPI, 96\% of tasks were completed correctly or with accurate problem identification.
With TAU/CUBE, the rate was 60\%.
Groups using TAU could not solve Task 1, as TAU lacks node-level representations.
These results were excluded.
Accuracy with EduMPI was similar across versions (A:~95\%, B:~98\%), but differed significantly with TAU/CUBE (A:~68\%, B:~52\%).
Individual tasks showed particular anomalies: Task~1 was answered correctly across the board, except TAU groups.
The color coding (T2) in EduMPI was understandable for all groups, while in CUBE, six out of nine groups expressed difficulties and requested a clearer legend.
Task~3 has only been answered correctly by four groups using CUBE or TAU.
Six groups stated that they did not understand the metrics presented.
Task~7, on the other hand, was solved correctly across the board in both tools.\\
\begin{figure}[H]
    \centering
    \vspace*{-10pt}

    \includegraphics[width=0.825\linewidth]{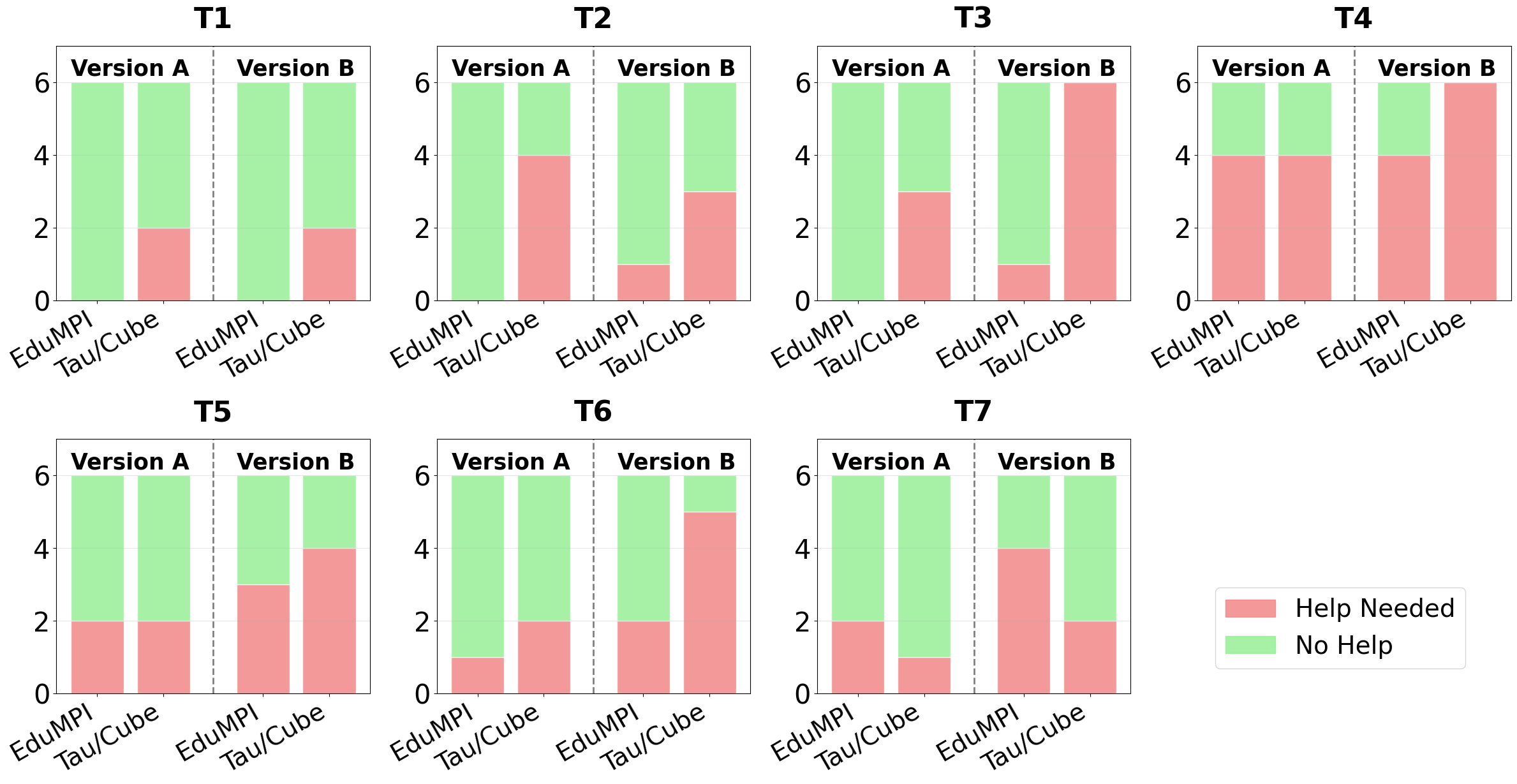}
    \captionof{figure}{Frequency of moderator assistance during task execution, compared across versions and tools}
    \label{fig:help}

    \vspace*{10pt}

    \includegraphics[width=0.85\linewidth]{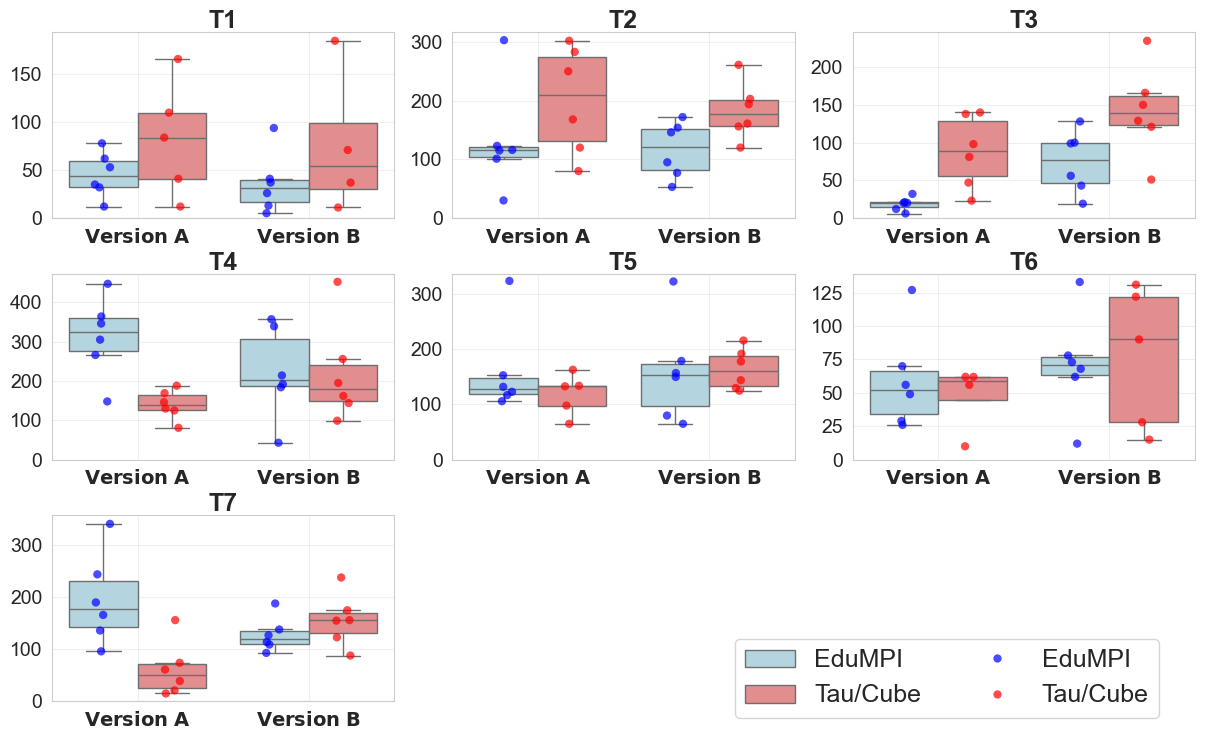}
    \captionof{figure}{Time spent per task compared across versions and tools}
    \label{fig:time}
\end{figure}
\textbf{Efficiency} was evaluated through the number of moderator interventions and the duration of task completion (see Figs.~\ref{fig:help} and~\ref{fig:time}).
Overall, the moderator only provided support when groups were completely stuck or had been pursuing a completely wrong approach for a long time.
Each run was time-limited by 90 minutes, and in both versions, emphasis was placed on being able to analyze as many tasks as possible correctly.
With EduMPI, interventions were required in 29\% of tasks, with TAU/CUBE, in 55\%.
With EduMPI, the number of interventions was lower in version A (21\%) than in version B (36\%). With TAU/CUBE, the difference was more pronounced (A:~43\%, B:~67\%). With the exception of Task~7, the groups consistently required more or equal support with TAU/CUBE than with EduMPI.
The processing times per task show a differentiated picture: EduMPI was consistently faster for tasks 1, 2, 3, 5, and 6.
For tasks 4 and 7, the groups required more time on average with EduMPI.
This was particularly noticeable in version A: Task 4 took twice as long as with TAU/CUBE. For Task 7, the groups needed more than three times as long on average.\\
\begin{figure}[H]
    \centering
    \vspace*{-10pt}
    \includegraphics[width=0.9\textwidth]{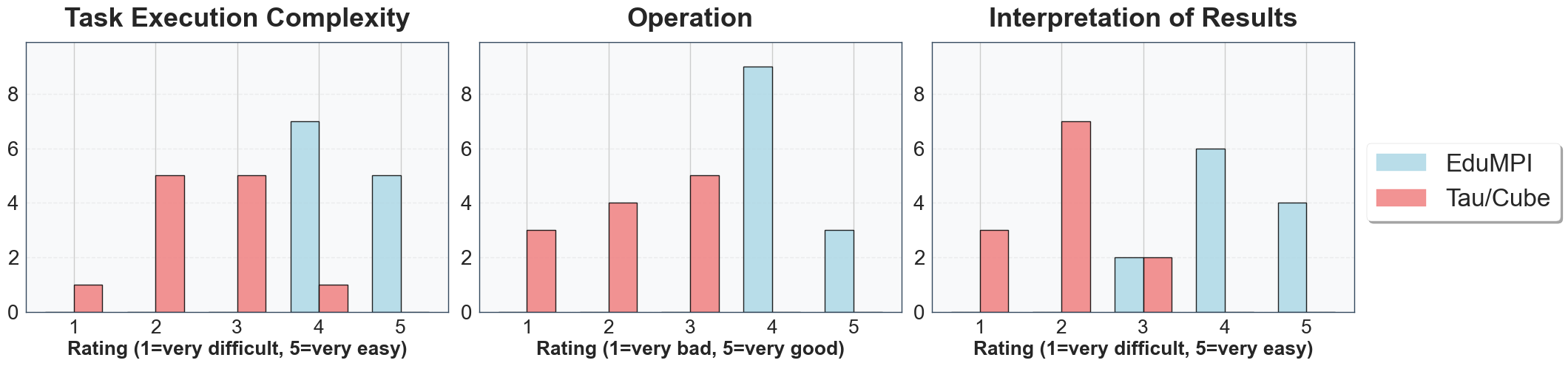} 
    \caption{Ratings based on group-averaged Likert responses regarding the tools}
    \label{fig:rating}
\end{figure}
\textbf{Satisfaction} was assessed through the groups' joint evaluation of the perceived difficulty, usability, and comprehensibility of the tools on a Likert scale.
In all categories, there was a clear preference for EduMPI (see Fig~\ref{fig:rating}).
The final questionnaires confirmed this trend: 24 of 33 participants stated that only EduMPI had improved their mental model of MPI program behavior, nine mentioned both tools, and none mentioned only TAU or CUBE.
When asked whether they would recommend using the tools in future courses, 32 students answered ``yes, definitely'' and one answered ``yes, with improvements'' for EduMPI.
For TAU/CUBE, the answers were: ``yes, definitely'' (1), ``yes, with improvements'' (14), ``not sure'' (14), ``probably not'' (4).
The 3D visualizations of performance data were used particularly frequently in TAU and EduMPI.
Participants who worked with CUBE repeatedly criticized the lack of a comparable view in their open feedback.

\section{Discussion}
Our usability study shows EduMPI significantly aids students in analyzing program behavior and identifying typical performance problems.
Building on prior work~\cite{RothJames2024} demonstrating reduced measurement barriers, this study reveals that students prefer EduMPI's visualizations over the professional tools TAU and CUBE.
Students consistently described EduMPI as more intuitive, accessible, and engaging than TAU/CUBE.
Although TAU/CUBE provide detailed performance metrics, their complexity and lack of exploratory support proved to be major obstacles.
Students required nearly twice as many interventions when using TAU/CUBE and frequently struggled with metric interpretation and navigation.
In contrast, 96\% of all tasks were solved correctly with EduMPI, compared to only 60\% using TAU/CUBE.
While our earlier study found that students often underestimated the value of performance analysis~\cite{RothJames2024}, the current results suggest a shift: all 33~participants recommended EduMPI for educational use.
This aligns with classroom experience, where EduMPI has fostered greater initiative, earlier cluster use, and a stronger focus on communication behavior and efficiency.
Overall, tools like EduMPI help shift attention from mere code execution to efficiency.\\
This study also supports the findings of~\cite{RothJamesKuhnFrisch2025} that understanding collective communication in MPI is highly valuable for students.
Five of six groups in version B (TAU/CUBE first) noted that process 0 finishes broadcasts faster, but assumed this was because it only sends data, while others receive, hence, sending is faster.
They did not consider the underlying algorithm, which neither TAU/CUBE makes visible.
In contrast, none of the version A groups (EduMPI first) made this assumption.
A key crossover result was a transfer effect: participants using EduMPI first performed better with TAU/CUBE, needing fewer interventions and completing more tasks.
Early EduMPI exposure thus facilitates transitioning to complex tools.
Therefore, EduMPI is not a replacement for professional tools but can serve as an effective bridge toward their later adoption.
Students in version A took over twice as long to complete Task 4 with EduMPI compared to TAU/CUBE, but answered it correctly every time (vs. 4 of 12 with TAU/CUBE).
This task was likely the most challenging, requiring the most moderator support.
The longer duration stemmed partly from students exploring all EduMPI views and options to identify both the bottleneck and the late broadcast issue.
In contrast, TAU/CUBE offered fewer analytical approaches to solve the task.
Task 7 (overall runtime analysis) revealed limitations in EduMPI's aggregation features.
Students found it easier to interpret time-segmented performance data but struggled to synthesize information across the entire runtime.
This likely results from EduMPI's real-time data structure, unlike the pre-aggregated profiles in TAU/CUBE.
Although global metrics are available in EduMPI, their interpretation was less intuitive, highlighting a potential area for improvement.

\section{Conclusion And Future Work}
EduMPI has proven to be an effective and easily accessible tool for supporting performance analysis in parallel programming education.
Our study shows that EduMPI facilitates entry into performance analysis and is recommended by students for use in teaching.
With EduMPI, more performance problems could be identified in a simple matrix multiplication than with the professional performance analysis tools CUBE/TAU, and students needed less support and usually less time for identification.
The study also showed that students who had previously analyzed the program with EduMPI could potentially achieve better results when analyzing with TAU and CUBE.
There is room for improvement in EduMPI in terms of visualizing aggregated data over the entire runtime.
In a further usability evaluation, we would also like to use EduMPI together with tracing-based tools.

\printbibliography
\end{document}